\documentclass[%
 reprint,
superscriptaddress,
 amsmath,amssymb,
 aps,
pra,
]{revtex4-2}
\usepackage{braket}
\usepackage{graphicx}
\usepackage{dcolumn}
\usepackage{bm}
\usepackage{hyperref}
\usepackage[caption=false]{subfig} 

\begin{document}

\preprint{APS/123-QED}

\title{Dynamical protection of quantum steering and fidelity dynamics in the double Jaynes–Cummings model}

\author{Koushik Mandal}
\email{koushikm@uwindsor.ca}
\affiliation{Department of Physics, University of Windsor, Windsor, Ontario, Canada}
\affiliation{Department of Physics and Computer Science, Wilfrid Laurier University, Waterloo, Ontario, Canada}

\author{Shohini Ghose}
\email{sghose@wlu.ca} 
\affiliation{Department of Physics and Computer Science, Wilfrid Laurier University, Waterloo, Ontario, Canada}

\author{Chitra Rangan}
\email{rangan@uwindsor.ca}
\affiliation{Department of Physics, University of Windsor, Windsor, Ontario, Canada}

\date{\today}

\begin{abstract}
We investigate the dynamics of Einstein–Podolsky–Rosen (EPR) steering in a double Jaynes–Cummings model, where two initially entangled spatially separated two-level atoms in two cavities interact with independent cavity modes. We study how the intrinsic noise in an initial Werner-type state affects the steering dynamics in this type of quantum optical system. We also analyze the evolution of steering under experimentally relevant conditions, including atom–cavity detuning and dipole–dipole interactions. We find that both detuning and dipole–dipole coupling help reduce steering sudden death in the system. We further identify a direct correlation between steering and state fidelity, revealing a threshold below which steering disappears. This suggests that fidelity may serve as a practical indicator of steerability in cavity QED systems. Our results provide insight into the controllability and robustness of non-classical correlations in realistic light–matter platforms.
\end{abstract}

\maketitle

\section{\label{sec:level1}Introduction}
Quantum correlations are a key feature of quantum information and quantum technologies such as quantum communication, cryptography\cite{PhysRevA.85.010301, Gehring2015, el2015quantum}, and metrology\cite{Huang2024Entanglement-enhanced, Liu2021Distributed, Shen2025Entanglement-enhanced}. Among these correlations, quantum entanglement has played a central role since the early days of research into quantum information and quantum computation, with its dynamical behavior being intensively investigated \cite{PhysRevA.67.029902, PhysRevA.70.052110, PhysRevLett.99.160502, PhysRevA.83.022109, xu2013experimental}. 

However, there is another kind of quantum correlation that occupies a unique and intermediate position in the hierarchy of non-classicality, lying strictly between entanglement and Bell nonlocality\cite{PhysRevLett.98.140402}. Originally introduced in the context of the Einstein-Podolsky-Rosen paradox\cite{PhysRev.47.777}, \emph{steering} captures the ability of one party to nonlocally affect the state of another through local measurements, under asymmetric trust assumptions.

Unlike entanglement, which is symmetric by definition, quantum steering is intrinsically directional. This directional nature makes steering particularly relevant for tasks such as one-sided device-independent(1SDI) quantum key distribution (QKD)\cite{PhysRevA.85.010301, el2015quantum} where only one measurement device must be trusted; subchannel communication\cite{PhysRevLett.114.060404}; and energy storage in quantum batteries\cite{https://doi.org/10.1002/qute.202500243}. As a result, understanding how steering behaves dynamically in realistic quantum systems is not only of foundational interest, but also of practical importance. Over the last two decades, EPR steering\cite{PhysRevLett.98.140402} has been the focus of a growing amount of research \cite{PhysRevA.80.032112, PhysRevLett.106.130402, PhysRevA.93.012108, PhysRevA.98.050104, PhysRevLett.113.140402, saunders2010experimental}.

The dynamical behavior of quantum steering remains comparatively unexplored in multipartite and cavity quantum optical systems. In particular, the phenomenon of Steering Sudden Death (SSD), where steerability vanishes abruptly at a finite time is interesting both from a theoretical and a practical perspective. SSD poses an obstacle for the reliable use of steering as a quantum resource. Identifying control mechanisms that suppress or eliminate SSD from the system is therefore of major interest.

The model we use to study the EPR steering dynamics is the Double Jaynes-Cummings Model (DJCM) consisting of two atoms placed in two cavities \cite{Yönaç_2006, PhysRevLett.97.140403, eberly2, Mandal_2024}. The DJCM is one of the paradigmatic models of quantum optics, and has been extensively used to study atom--field coherence, as well as the generation, transfer, collapse-revival phenomena, and sudden death effects in entanglement\cite{Pandit_2018, li2020entanglement, laha2023dynamics, jakubczyk2017quantum, PhysRevA.76.042313}. This model provides a natural and experimentally feasible platform to investigate the steering dynamics of quantum optical systems. While entanglement dynamics in this model is well understood, there are few investigations of quantum steering dynamics in the DJCM. In \cite{ban2025steerability}, the authors studied the the steerability dynamics of two types of Bell-like states in double Jaynes–Cummings model under noiseless and noisy environments. In another study, Khouja et al.\cite{Khouja_2023} investigated the quantum steering of two Tavis-Cummings atoms with dipole-dipole interaction under intrinsic decoherence. Both these studies focused on initially pure states of atoms.

In experimental situations, pure states are difficult to prepare and maintain due to unavoidable environmental noise and imperfect state preparation. For this reason, mixed states provide a more realistic starting point for dynamical studies. In this scenario, Werner-type states are very suitable since these states contain noise in their structure naturally. It was Werner\cite{PhysRevA.40.4277} who first showed that there are states (now known as Werner states) that show entanglement but not Bell nonlocality. 
Later, it was shown that not all entangled Werner states are steerable \cite{PhysRevLett.98.140402}, and recent studies have further clarified their steerability limits under general measurements \cite{PhysRevLett.132.250202,PhysRevLett.132.250201}, namely, Werner states with any visibility $r \le 1/2$ under general POVMs are not steerable.  In addition, Werner-type states have been prepared experimentally and widely employed in studies of teleportation, noisy quantum channels, and correlation quantification \cite{PhysRevLett.92.177901,PhysRevLett.84.4236, Czerwinski_2021}.

This work is motivated by the following questions:
\begin{itemize}
    \item How does quantum steering between two atoms evolve in the DJCM when the initial atomic state is in a mixed state?

    \item Is there a method to detect exactly when steering disappears and revives?
    
    \item Can control parameters such as detuning and dipole-dipole coupling strength affect steering sudden death?
    
\end{itemize}

To address these questions, we investigate the dynamics of atomic quantum steering starting from Werner-type initial atomic states, with both cavities initially in the vacuum state. By combining analytical treatment with numerical simulations, we demonstrate that interaction-induced coherence can completely remove steering sudden death for visibility parameters exceeding a critical threshold. 

The significance of our work lies in three key aspects.  First, we characterize the dynamical behavior of quantum steering under different initial Werner-type mixed atomic states (going beyond the Bell-like pure-state analysis of \cite{ban2025steerability} in DJCM systems). We demonstrate that quantum steering is fundamentally more fragile than entanglement under local cavity interactions; specifically, the atomic subsystem experiences steering sudden death (SSD) during which the two-atom state remains entangled, but completely loses  operational steerability. Second, we establish a direct connection between steering and quantum state fidelity dynamics. We show that the `cusps'-like features in the maximized fidelity envelopes act as highly-sensitive predictive indicators for the precise temporal thresholds where the state enters or exits the Local Hidden State (LHS) manifold. 

Finally, we demonstrate that SSD can be actively manipulated through control parameters. By introducing atom-cavity detuning ($\Delta$) and interatomic Dipole-Dipole Interaction (DDI), we demonstrate how these parameters stabilize atomic correlations and shield the state from cavity-induced loss. Our results advance the understanding of quantum steering as a dynamical resource in quantum systems and identify concrete physical mechanisms for its protection and control. These findings are expected to be relevant for future experiments in cavity and circuit QED platforms, as well as for the development of steering-based quantum technologies.

The remainder of this paper is organized as follows. 
In Sec.~\ref{sec2}, we present the theoretical framework 
of the double Jaynes-Cummings model, specifying the system  Hamiltonian, the initial Werner state of the  atomic subsystem, and the vacuum configuration of the quantized radiation fields. We further derive the explicit conditions and thresholds governing EPR steering within  this architecture. In Sec.~\ref{sec3}, we analyze the comparative dynamical evolution of bipartite entanglement and quantum steering in the non-interacting regime, and elucidate the underlying physical mechanisms through the lens of quantum state fidelity. Sections~\ref{sec4} and \ref{sec5} extend this analysis to more experimentally relevant regimes, examining in turn the effects of atom-cavity detuning and dipole-dipole interactions on the robustness, persistence, and directional asymmetry of the steering dynamics.

\section{Theoretical framework}
\label{sec2}

\subsection{Quantum Steering}

Quantum steering\cite{PhysRev.47.777, PhysRevLett.98.140402, PhysRevA.80.032112, Lee2025} describes the ability of one observer to nonlocally affect the set of conditional states of another distant observer through local measurements. Consider a bipartite quantum state represented by the density operator $\hat{\rho}_{AB}$ shared between two parties, Alice and Bob. The state is said to be unsteerable from Alice to Bob if the joint probability distribution can be expressed in the Local Hidden-State (LHS) model
\begin{equation}
P(a,b|x,y)
=
\sum_{\lambda}
P(\lambda)\,
P(a|x,\lambda)\,
P_Q(b|y,\rho_\lambda),
\end{equation}
where $x$ and $y$ denote the measurement settings chosen by Alice and Bob, respectively, while $a$ and $b$ represent their corresponding measurement outcomes. Here, $\lambda$ is a hidden variable distributed according to the classical probability distribution $P(\lambda)$, and
\begin{equation}
P_Q(b|y,\rho_\lambda)
=
\mathrm{Tr}\left[M_{b|y}\rho_\lambda\right]
\end{equation}
is the quantum probability distribution associated with Bob's local quantum state $\rho_\lambda$, with $M_{b|y}$ denoting Bob's measurement operator.

The existence of such a decomposition implies that Bob's subsystem can be completely described by a pre-existing ensemble of local quantum states, independent of any nonclassical influence from Alice's measurements. Therefore, violation of the LHS model certifies the presence of EPR steering. Unlike entanglement, steering is inherently directional, meaning that steering from Alice to Bob and from Bob to Alice are generally not equivalent.

\subsection{Double Jaynes--Cummings model (DJCM)}

\begin{figure}[ht!]
\centering
\includegraphics[width=\linewidth]{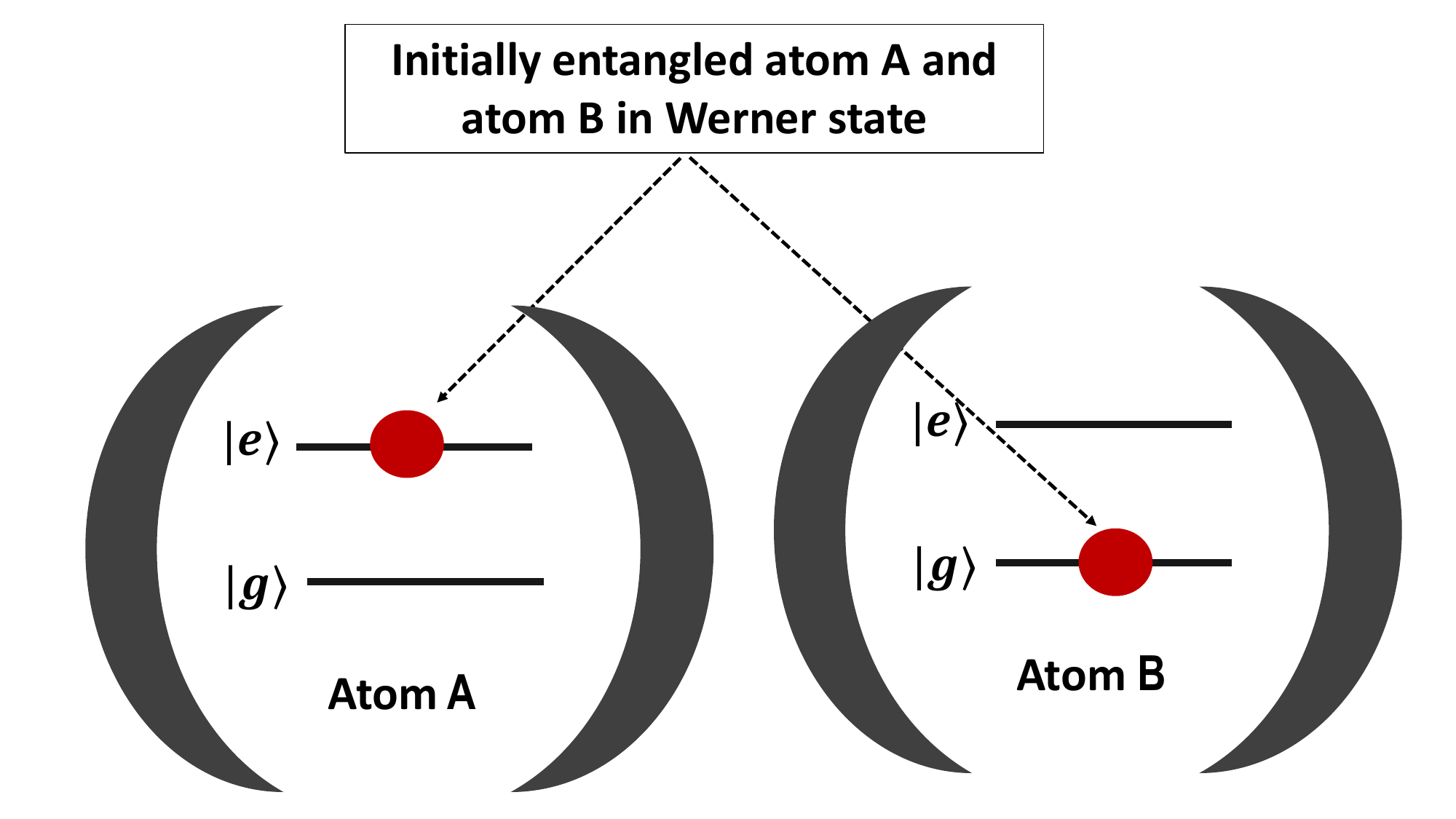}
\caption{Schematic representation of the double Jaynes--Cummings model consisting of two spatially separated two-level atoms, each interacting locally with an independent cavity mode.}
\label{fig:djcm}
\end{figure}

We consider a double Jaynes-Cummings model (DJCM) composed of two non-interacting two-level atoms, each locally coupled to an independent single-mode cavity field, as illustrated in Fig.~\ref{fig:djcm}. Such systems provide an important platform for studying the dynamics of quantum correlations in cavity-QED and light-matter interaction systems. Throughout this work, we use units in which $\hbar = 1$.

Under the rotating-wave approximation, the total Hamiltonian of the DJCM system can be written as
\begin{align}
\hat{H}_{\text{tot}}
&=
\frac{\omega_{0}}{2}\,\hat{\sigma}_{z}^{A}
+
\frac{\omega_{0}}{2}\,\hat{\sigma}_{z}^{B}
+
g\left(
\hat{a}^{\dagger}\hat{\sigma}_{-}^{A}
+
\hat{a}\hat{\sigma}_{+}^{A}
\right)
\nonumber\\
&\quad
+
g\left(
\hat{b}^{\dagger}\hat{\sigma}_{-}^{B}
+
\hat{b}\hat{\sigma}_{+}^{B}
\right)
+
\nu\,\hat{a}^{\dagger}\hat{a}
+
\nu\,\hat{b}^{\dagger}\hat{b},
\label{djcmmodel}
\end{align}
where $\omega_{0}$ and $\nu$ denote the atomic transition frequency and cavity-field frequency, respectively. The operators $\hat{\sigma}_{z}^{i}$ $(i=A,B)$ are the usual Pauli operators for the two-level atoms, while $\hat{\sigma}_{+}^{i}$ and $\hat{\sigma}_{-}^{i}$ represent the corresponding raising and lowering operators. The operators $\hat{a}$ $(\hat{a}^{\dagger})$ and $\hat{b}$ $(\hat{b}^{\dagger})$ are the annihilation (creation) operators associated with the two cavity modes. The parameter $g$ characterizes the strength of the local atom--field interaction, taken as the same for both sets of atom+cavity. 

In the present model, the two atoms do not interact directly with each other. As a result, all quantum correlations between the atoms originate either from the initially prepared atomic state or from the redistribution of correlations induced by the local atom--field interactions during the evolution. The total Hilbert space of the composite system is
\begin{equation}
\mathcal{H}
=
\mathcal{H}_{A}
\otimes
\mathcal{H}_{B}
\otimes
\mathcal{H}_{a}
\otimes
\mathcal{H}_{b},
\end{equation}
where $\mathcal{H}_{A}$ and $\mathcal{H}_{B}$ correspond to the atomic subspaces, while $\mathcal{H}_{a}$ and $\mathcal{H}_{b}$ represent the Hilbert spaces of the two cavity modes.

The basis states of the total system are written as
\begin{equation}
\ket{\text{Atom A}}
\otimes
\ket{\text{Atom B}}
\otimes
\ket{\text{Cavity a}}
\otimes
\ket{\text{Cavity b}},
\end{equation}
where the atomic states are denoted by $\ket{g}$ and $\ket{e}$ corresponding to the ground and excited states, respectively, while the cavity modes are described by the photon-number states $\ket{n}$.

\subsection{Werner-Type initial atomic states}

Initially, the two atoms are assumed to be prepared in a Werner-type mixed entangled state~\cite{PhysRevA.40.4277}
\begin{equation}
\hat{\rho}_{W}(0)
=
p\ket{\Psi^{-}}\bra{\Psi^{-}}
+
\frac{1-p}{4}\hat{\mathbb{I}}_{4},
\end{equation}
where
\begin{equation}
\ket{\Psi^{-}}
=
\frac{1}{\sqrt{2}}
\left(
\ket{eg}
-
\ket{ge}
\right),
\end{equation}
and $p$ $(0 \leq p \leq 1)$ is the visibility parameter characterizing the weight of the maximally entangled component in the Werner state.

Werner states exhibit a well-known hierarchy of quantum correlations: they are separable for $p \leq 1/3$, entangled but Bell-local for $1/3 < p \leq 1/\sqrt{2}$, and violate Bell inequalities only for sufficiently large $p$. Quantum steering occupies an intermediate regime within this hierarchy. Therefore it is a sensitive probe of nonclassical correlations in open quantum systems.

\subsection{Reduced density matrix for the atomic subsystem}

Let $\hat{\rho}_{\mathrm{tot}}(0)$ denote the initial state of the full system,
\begin{equation}
\hat{\rho}_{\mathrm{tot}}(0)
=
\hat{\rho}_{AB}(0)
\otimes
\hat{\rho}_{a}(0)
\otimes
\hat{\rho}_{b}(0),
\end{equation}
where $\hat{\rho}_{a,b}(0)$ represents the initial state of cavities $a$ and $b$ respectively. For our systems these are vacuum states. So, the total initial state is
\begin{equation}
\hat{\rho}_{\text{tot}}(0)
=
\hat{\rho}_{AB}(0)
\otimes
\ket{00}\bra{00}.
\end{equation}

In the computational basis
$\{\ket{ee},\ket{eg},\ket{ge},\ket{gg}\}$, the initial density matrix takes the form
\begin{equation}
\hat{\rho}_{AB}(p)
=
\begin{pmatrix}
\frac{1-p}{4} & 0 & 0 & 0 \\[6pt]
0 & \frac{1+p}{4} & -\frac{p}{2} & 0 \\[6pt]
0 & -\frac{p}{2} & \frac{1+p}{4} & 0 \\[6pt]
0 & 0 & 0 & \frac{1-p}{4}
\end{pmatrix}.
\end{equation}

The time evolution of the total system is governed by the unitary operator
\begin{equation}
\hat{U}(t)
=
e^{-i\hat{H}_{\text{tot}}t},
\end{equation}
such that the evolved density operator becomes
\begin{equation}
\hat{\rho}_{\text{tot}}(t)
=
\hat{U}(t)\,
\hat{\rho}_{\text{tot}}(0)\,
\hat{U}^{\dagger}(t).
\end{equation}
Appendix \ref{app_A} shows that the reduced two-atom density matrix as a function of time can be calculated analytically.
The resulting state preserves the X-state structure throughout the evolution, which considerably simplifies the analysis of entanglement, steering, and fidelity dynamics. In the basis
$\{\ket{ee},\ket{eg},\ket{ge},\ket{gg}\}$, the reduced density matrix is given by
\begin{equation}
\hat{\rho}_{AB}(t)
=
\begin{pmatrix}
\rho_{11} & 0 & 0 & 0 \\[6pt]
0 & \rho_{22} & \rho_{23} & 0 \\[6pt]
0 & \rho_{32} & \rho_{33} & 0 \\[6pt]
0 & 0 & 0 & \rho_{44}
\end{pmatrix},
\end{equation}
where the nonvanishing matrix elements are
\begin{align}
\rho_{11}
&=
\frac{1-p}{4}C^4,
\\
\rho_{22}
=
\rho_{33}
&=
\frac{C^2}{4}
\left[
2p
+
(1-p)(1+S^2)
\right],
\\
\rho_{23}
=
\rho_{32}
&=
-\frac{p}{2}C^2,
\\
\rho_{44}
&=
pS^2
+
\frac{1-p}{4}(1+S^2)^2.
\end{align}

\subsection{Measures of EPR steering: CJWR steering inequality}

To study the steering dynamics of the atomic subsystem, we use the three-setting Cavalcanti--Jones--Wiseman--Reid (CJWR)\cite{PhysRevA.80.032112} steering inequality, which is widely used for detecting EPR steering in two-qubit systems. In this framework, Alice performs measurements along three mutually orthogonal Pauli directions, namely $\hat{\sigma}_x^{A}$, $\hat{\sigma}_y^{A}$, and $\hat{\sigma}_z^{A}$, while Bob performs the corresponding trusted Pauli measurements on his subsystem.

The correlation functions are defined as
\begin{equation}
T_{jk}(t)
=
\mathrm{Tr}
\left[
\left(
\hat{\sigma}_j^{A} \otimes \hat{\sigma}_k^{B}
\right)
\hat{\rho}_{AB}(t)
\right],
\qquad
j,k \in \{x,y,z\},
\end{equation}
where $\rho_{AB}(t)$ is the reduced density matrix of the two-atom subsystem at time $t$.

Using these correlations, the steering parameter from Alice to Bob can be written as
\begin{equation}
S_{A\rightarrow B}(t)
=
\sqrt{
T_{xx}^2(t)
+
T_{yy}^2(t)
+
T_{zz}^2(t)
}.
\label{eq:cjwr}
\end{equation}

The state is said to be steerable from Alice to Bob whenever
\begin{equation}
S_{A\rightarrow B}(t) > 1.
\end{equation}

In the present work, this quantity is used to analyze the dynamical behavior of steering and the appearance of steering sudden death (SSD) in the atom--atom subsystem.

\subsection{Measure of entanglement}

To quantify the entanglement between the two atoms, we use concurrence, introduced by Wootters~\cite{wootters2001entanglement}. For a general two-qubit density matrix $\hat{\rho}_{AB}$, the concurrence is defined as
\begin{equation}
C_{AB}
=
\max
\left\{
0,
\Lambda_1-\Lambda_2-\Lambda_3-\Lambda_4
\right\},
\end{equation}
where $\Lambda_i$ $(i=1,2,3,4)$ are the square roots of the eigenvalues, arranged in decreasing order, of the matrix
\begin{equation}
\hat{R}
=
\hat{\rho}_{AB}
\left(
\hat{\sigma}_{y}^{A}\otimes \hat{\sigma}_{y}^{B}
\right)
\hat{\rho}_{AB}^{*}
\left(
\hat{\sigma}_{y}^{A}\otimes \hat{\sigma}_{y}^{B}
\right).
\end{equation}

The concurrence takes values in the range $0 \leq C_{AB} \leq 1$, where $C_{AB}=0$ corresponds to a separable state and $C_{AB}=1$ represents a maximally entangled state. Since steering is a stronger form of quantum correlation than entanglement, comparing the dynamics of concurrence and steering allows us to identify regimes where entanglement survives even in the absence of steerability.

\section{EPR steering and entanglement for non-interacting Hamiltonian}
\label{sec3}

\subsection{Comparison between quantum steering and entanglement}
\begin{figure}
    \centering
\includegraphics[width=\linewidth]{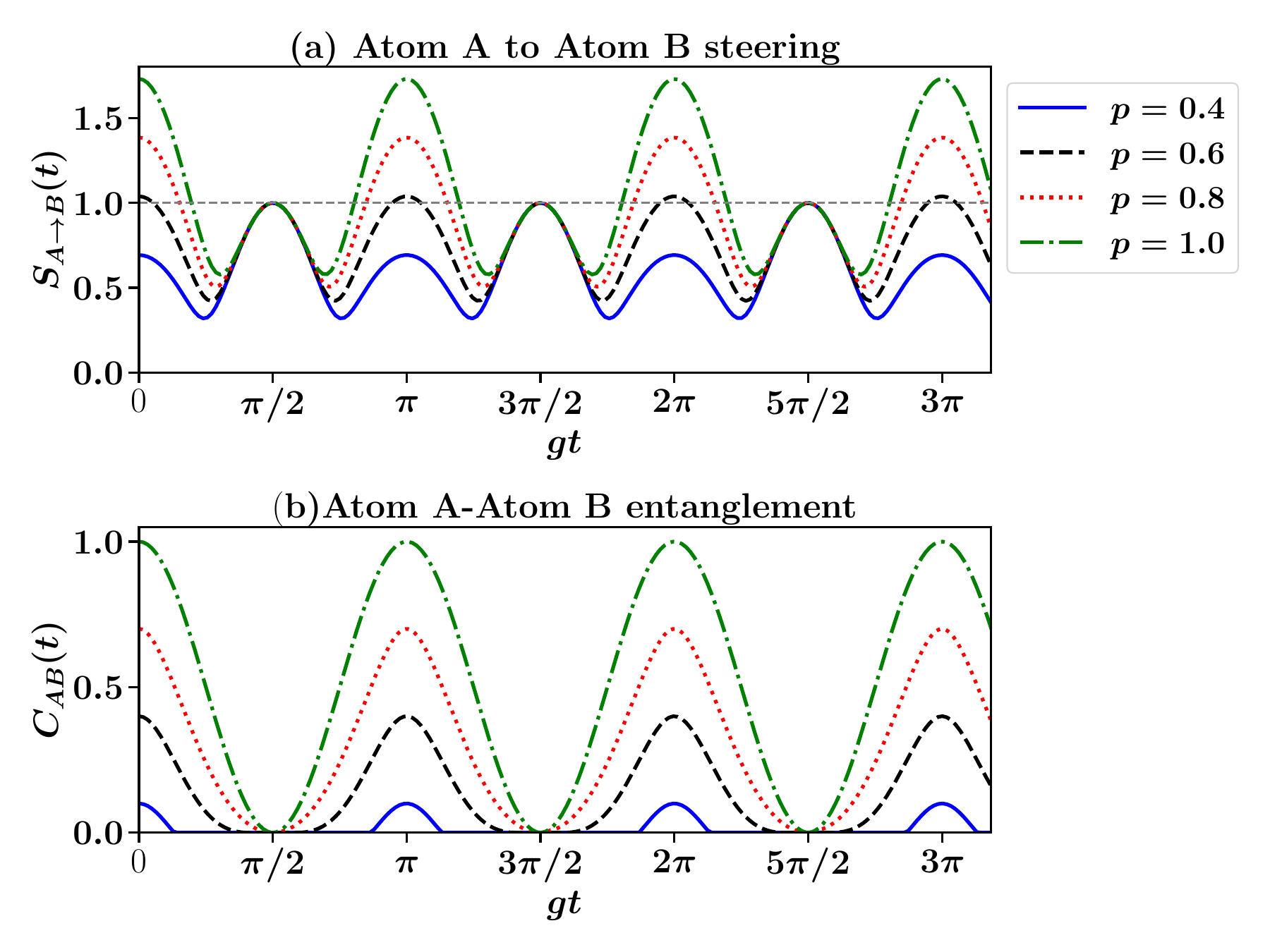}
 \caption{Dynamics of (a) quantum steering from atom A to atom B and (b) atom--atom entanglement for different values of the visibility parameter $p$. The chosen values $p = 0.4, 0.6, 0.8,$ and $1.0$ correspond to initial Werner states ranging from mixed entangled states to the maximally entangled Bell state.}
    \label{fig:steer_conc}
\end{figure}

In this section, we compare bipartite entanglement and quantum steering for the two-atom subsystem in the double Jaynes--Cummings model. Although both correlations originate from the same underlying quantum coherence, they exhibit markedly different dynamical behaviors and robustness properties. Both of these dynamics are presented in Fig. \ref{fig:steer_conc}.

Entanglement between the atoms is quantified using the concurrence $C(t)$, while steerability is characterized by the CJWR-type steering parameter $S_{A\to B}(t)$. For the initial Werner state, entanglement is present whenever the visibility parameter satisfies $p > 1/3$, whereas steering requires the stricter condition $p > 1/2$. This immediately establishes a hierarchy between the two forms of quantum correlations. 

The periodic oscillations observed in both $C(t)$ and $S_{A\to B}(t)$ are a direct manifestation of coherent Rabi-like oscillations. As time evolves, the initial electronic excitation and quantum correlations stored within the atomic subsystem are dynamically transferred to other atom-cavity and cavity-cavity subsystems, and subsequently mapped back onto the atoms.

The structural evolution of these curves in Fig. \ref{fig:steer_conc} highlights that quantum steering is fundamentally more fragile than entanglement under local cavity-atom interactions. For moderate state visibilities (e.g., $p = 0.6$ and $p = 0.4$), the concurrence $C(t)$ touches zero only periodically at the half-cycle points ($gt = \pi/2, 3\pi/2, \dots$), representing standard entanglement swapping with the other subsystems. Conversely, the steering parameter $S_{A\to B}(t)$ drops below the critical classical threshold ($S \le 1$, marked by the dashed line) much earlier and remains dead for extended finite durations. This introduces a pronounced ``steering sudden death'' window where the state remains entangled but completely loses its steerability. This clearly demonstrates that entanglement alone is insufficient to guarantee steerability. The rapid collapse of $S_{A\to B}(t)$ implies that as the atoms interact with their local cavity modes, the local measurement back-action on Atom A rapidly loses its capacity to non-classically steer the state of Atom B. The cavity acts as a local information sink that disrupts the steering mechanism long before it fully dismantles the underlying atomic entanglement.

\subsection{Fidelity of the time-evolved state with Werner states}
We hypothesize that the SSD in the system is associated with the decrease in the overlap of the time-evolved state with the initial Werner states.  Therefore, we calculate the quantum state fidelity $F(t)$ between the time-evolved reduced density matrix $\hat{\rho}_{AB}(t)$ and the initial atomic Werner state $\hat{\rho}_W(0)$, which is defined by the standard Uhlmann formula:
\begin{equation}
F(t) = \left[ \text{Tr} \sqrt{\sqrt{\hat{\rho}_W(0)} \hat{\rho}_{AB}(t) \sqrt{\hat{\rho}_W(0)}} \right]^2.
\label{eq:fidelity_def}
\end{equation}
Because both the initial Werner state and the time-evolved density matrix retain a block-diagonal X-structure inside the computational basis $\{|ee\rangle, |eg\rangle, |ge\rangle, |gg\rangle\}$, Eq.~\eqref{eq:fidelity_def} can be simplified analytically by decoupling the system into distinct subspaces (the detailed algebraic derivation is provided in Appendix B). This formulation allows us to map the loss of steerability directly onto the degradation of state overlap as the open system evolves.
\begin{figure*}[t] 
    \centering
    \subfloat[]{\label{fidel_dyna_a}\includegraphics[width=0.5\textwidth]{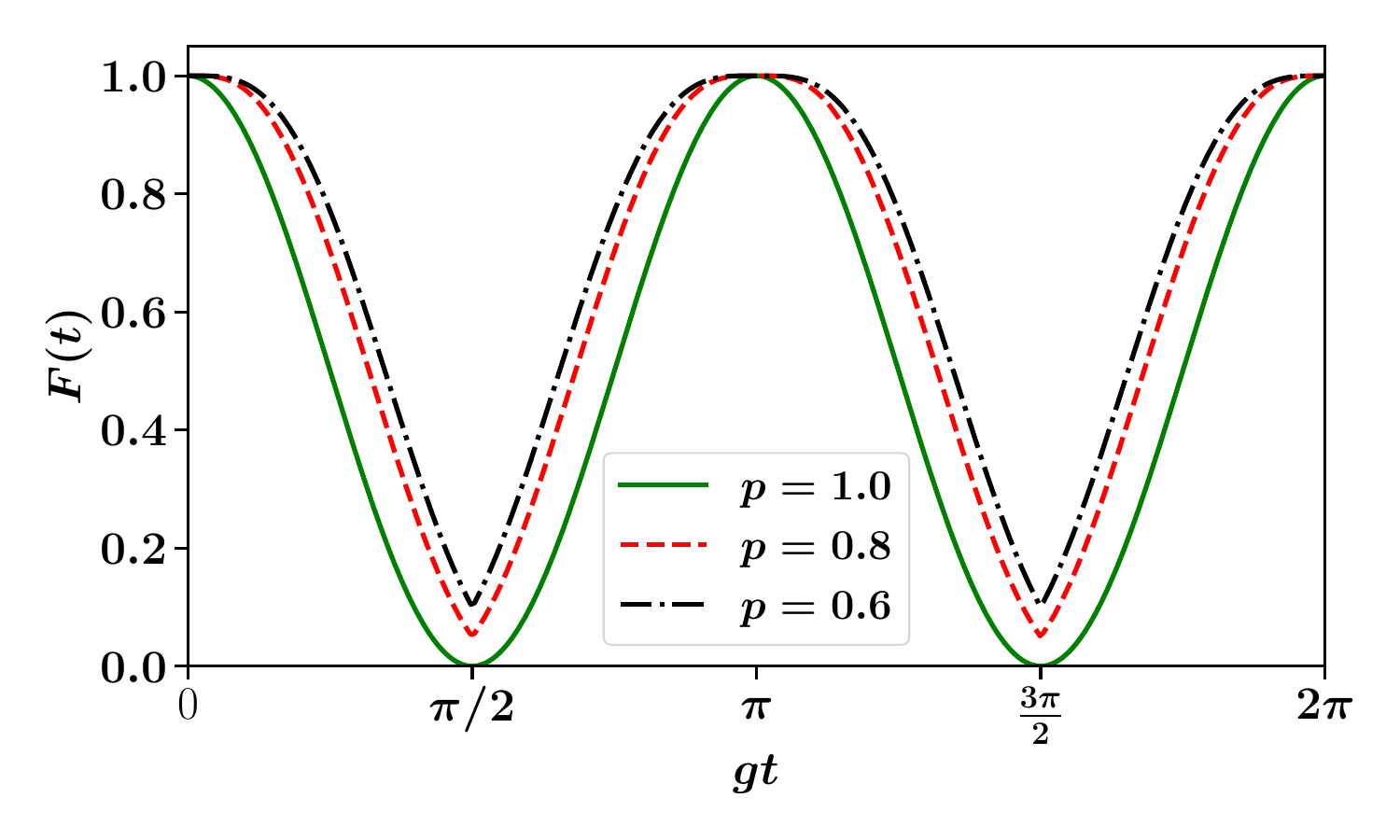}}
    \hfill
    \subfloat[]{\label{fidel_dyna_b}\includegraphics[width=0.48\textwidth]{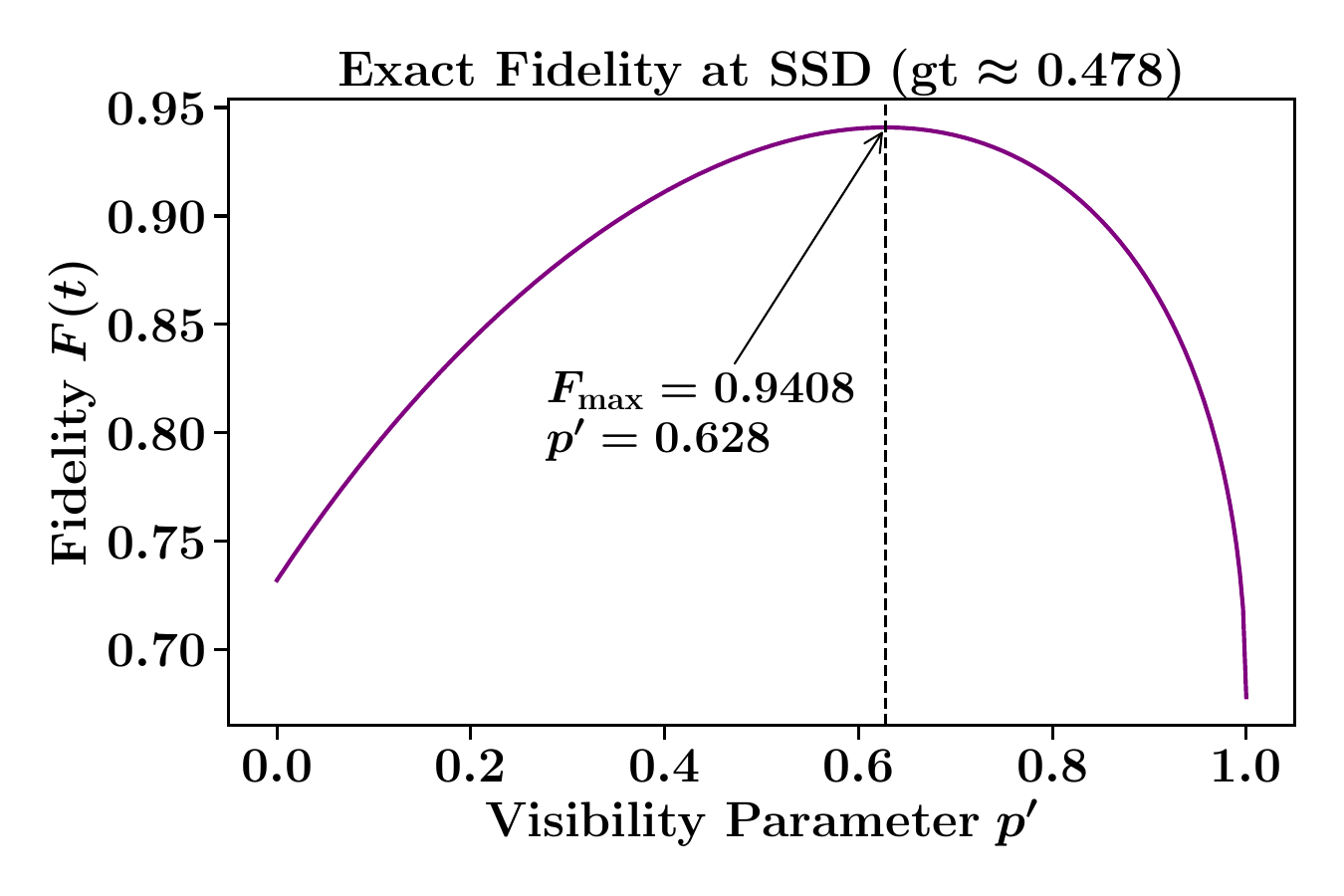}}
    
    \vspace{0.4cm} 
    
    \subfloat[]{\label{fidel_dyna_c}\includegraphics[width=0.5\textwidth]{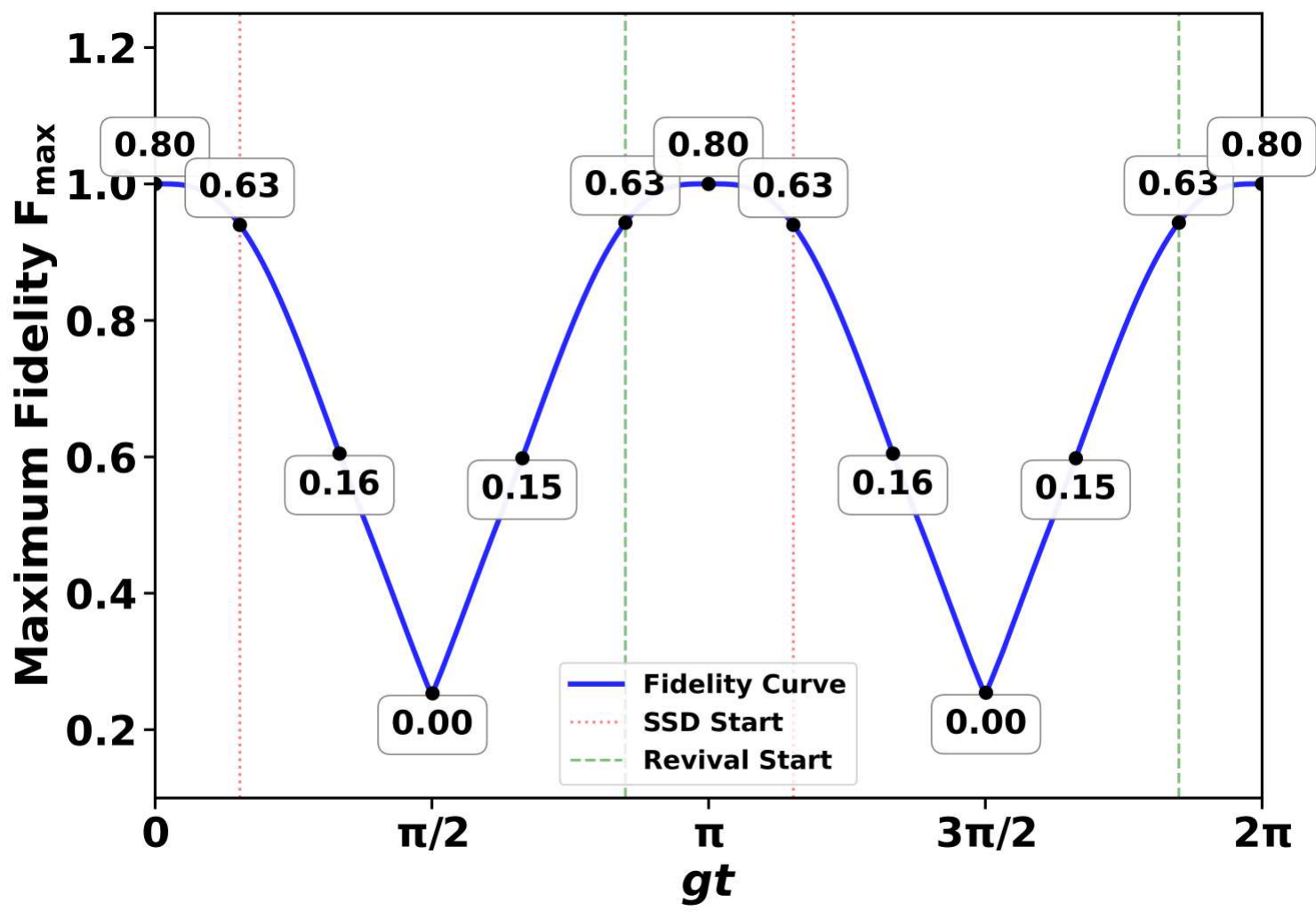}}

    \caption{Panel (a) shows the time evolution of the fidelity for different initial Werner-state parameters $p$. In panel (b), we plot the fidelity between the evolved atomic state and the entire family of Werner states at the onset of steering sudden death for the initial condition $p=0.8$. Panel (c) illustrates how the evolved state overlaps with different Werner states at several time instants for the same initial state. The numbers displayed inside the boxes denote the value of the Werner parameter $p'$ corresponding to the maximum fidelity.}
    \label{fidel_dyna}
\end{figure*}

Figure \ref{fidel_dyna_a} illustrates the time evolution of the fidelity $F(t)$ for different initial visibility parameters $p$. As the atom-cavity interaction progresses, the fidelity between the evolved atomic state and its initial configuration monotonically decreases. This decay signals that the local atom-field coupling is systematically redirecting quantum coherence away from the atomic subsystem and into the cavity modes. Within these specific temporal windows, the remaining atomic correlations drop below a critical threshold where the reduced state admits a local hidden state (LHS) description, thereby precipitating the onset of steering sudden death. Conversely, when the coherence is transferred back into the atomic degrees of freedom, the fidelity revives, the LHS description becomes invalid, and EPR steerability is restored.

To uncover this mechanism in greater detail, we evaluate how the evolved state traverses through the entire geometric space of the Werner state family. Let $\hat{\rho}_W(p')$ characterize a target Werner state parameterized by an arbitrary visibility parameter $p'$. We define the maximum fidelity as:
\begin{equation}
F_{\max}(t) = \max_{p'} F\big(\hat{\rho}_{AB}(t), \hat{\rho}_W(p')\big).
\label{eq:max_fidelity}
\end{equation}

In Fig.~\ref{fidel_dyna_b}, we chart the profile of this fidelity optimization at the precise onset of steering sudden death ($gt \approx 0.478$) for an initial state of $p = 0.8$. At this threshold, the state exhibits a maximum fidelity of $F_{\max} \approx 94\%$ at an optimal parameter value of $p' = 0.63$. This demonstrates that while the state has deformed away from its initial state, it still structurally mimics a highly entangled Werner state.

\begin{table}[!htbp]
\caption{\label{tab:werner_fidelity}Maximum fidelity between the time-evolved atomic states and the family of Werner states at different time instants. The corresponding Werner-state parameter associated with the highest fidelity is also shown. The initial atomic state is chosen with visibility parameter $p=0.8$.}
\begin{center} 
\setlength{\tabcolsep}{12pt} 
\begin{tabular}{ccc}
\hline\hline 
Time $gt$ & $F_{\max}$ & $p'$ \\
\hline
0.00 & 1.00 & 0.80 \\
0.49 & 0.93 & 0.63 \\
1.03 & 0.61 & 0.16 \\
1.56 & 0.25 & 0.00 \\
2.07 & 0.59 & 0.15 \\
2.68 & 0.94 & 0.63 \\
3.13 & 1.00 & 0.80 \\
3.62 & 0.93 & 0.62 \\
4.19 & 0.60 & 0.15 \\
4.70 & 0.25 & 0.00 \\
5.21 & 0.58 & 0.13 \\
5.81 & 0.94 & 0.63 \\
6.28 & 1.00 & 0.80 \\
6.76 & 0.94 & 0.62 \\
7.32 & 0.60 & 0.15 \\
7.85 & 0.25 & 0.00 \\
8.36 & 0.59 & 0.14 \\
8.95 & 0.94 & 0.63 \\
9.42 & 1.00 & 0.80 \\
\hline\hline 
\end{tabular}
\end{center}
\end{table}

However, as the evolution continues toward the half-cycle point ($gt = \pi/2$), the maximum fidelity drops precipitously to approximately $25\%$, as depicted in Fig.~\ref{fidel_dyna_c} and tabulated in Table 1. This severe reduction occurs because the initial atomic excitations have completely mapped into the photonic fields, leaving the atoms in a near-vacuum state $|gg\rangle$. This structural shift drags the state far away from the ``steerable'' domain of the Werner family, creating a regime dominated by trivial classical correlations that cannot support non-local steering. Following this minimum, $F_{\max}(t)$ recovers periodically, establishing a clear dynamical link between the structural fidelity of the state and the birth-and-death windows of SSD.

An interesting feature of the optimized fidelity curves shown in Fig.~\ref{fidel_dyna_c} is the appearance of sharp, non-analytic `cusps'-like features near the collapse-revival boundaries. These sharp changes are not due to any actual discontinuity in the system dynamics, since the evolution of the quantum state itself remains continuous. Rather, these features arise from the optimization procedure over the family of Werner states used to define the maximum fidelity. Around these regions, the evolved atomic state rapidly shifts between different optimal Werner-like structures, producing cusp-like features in the fidelity dynamics. Physically, this reflects a rapid redistribution and restructuring of quantum correlations during the collapse and revival process. This kind of behaviour has been observed before for other quantum correlations, e.g., in quantum discord~\cite{PhysRevLett.107.140403}. 

In practical cavity-QED architectures, where full state tomography to verify steering can be resource-intensive, monitoring these specific non-analyticities in the fidelity provides a high-sensitivity signature for the birth or death of steerable correlations.

\section{Quantum steering dynamics in the presence of detuning}
\label{sec4}

Here, we investigate the influence of atom--cavity detuning on the quantum steering between the two atoms. 
In this kind of system, detuning can work as a control knob. For simplicity, the two atoms are detuned in opposite ways, i.e., atom $A$ is blue-detuned and atom $B$ is red-detuned by the same magnitude.
The Hamiltonian with detuning can be written as

\begin{equation}
\hat{H}_{\text{int}} = \Delta \sum_{j \in \{A, B\}} \hat{\sigma}_{z}^{j} + g \left[ \left( \hat{a}^{\dagger}\hat{\sigma}_{-}^{A} + \hat{b}^{\dagger}\hat{\sigma}_{-}^{B} \right) + \text{H.c.} \right].
\end{equation}
The effective Hamiltonian matrix is 
\begin{equation}
H_j =
\begin{pmatrix}
\Delta_j/2 & g \\
g & -\Delta_j/2
\end{pmatrix}.
\end{equation}
Its eigenfrequencies are
\begin{equation}
\Omega_j = \sqrt{g^2 + (\Delta_j/2)^2}.
\end{equation}

If the atom starts excited and the cavity is in a vacuum state then:
\begin{equation}
|e,0\rangle_j \;\longrightarrow\;
\alpha_j(t)|e,0\rangle_j + \beta_j(t)|g,1\rangle_j
\end{equation}
with
\begin{align}
\alpha_j(t) &= 
\cos(\Omega_j t)
- i\frac{\Delta_j}{2\Omega_j}\sin(\Omega_j t), \\
\beta_j(t) &=
- i\frac{g}{\Omega_j}\sin(\Omega_j t).
\end{align}

\begin{figure}[ht!]
\centering
\includegraphics[width=\linewidth]{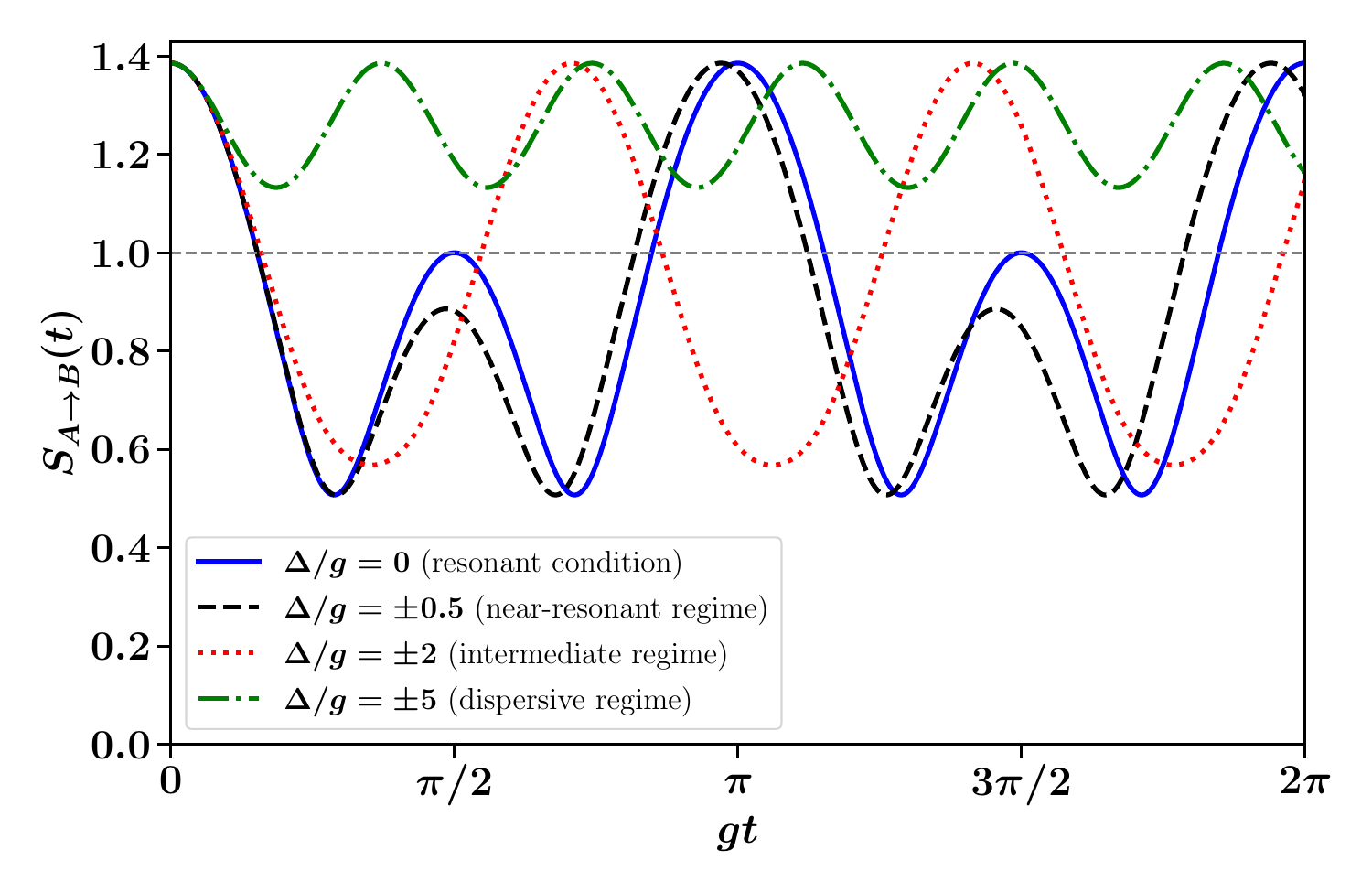}
\caption{Dynamics of quantum steering in the presence of atom--cavity detuning for the initial Werner-state parameter $p=0.8$. The steering dynamics are shown for different detuning regimes, including the \textit{resonant, near-resonant, intermediate, and dispersive} regimes. Positive and negative values of detuning $(\pm\Delta)$ correspond to blue- and red-detuned atomic transitions for atoms $A$ and $B$, respectively.}
\label{detuned_steering}
\end{figure}

The effects of detuning are studied in different regimes of interactions, namely in the \textit{resonant, near-resonant, intermediate and dispersive} regions. Figure \ref{detuned_steering} shows the temporal behavior of the steering 
$S_{A\rightarrow B}(t)$ for increasing values of detuning $\Delta$, with both cavity modes initially prepared in the vacuum state. In the resonant case $\Delta=0$, the system exhibits pronounced steering sudden death, characterized by finite temporal intervals during which the steering inequality is no longer violated. This behavior originates from the coherent exchange of excitation between each atom and its local cavity mode, which periodically depletes atomic coherence and makes the reduced two-atom state
insufficiently steerable.

However, introducing a small fractional detuning ($\Delta=0.5$) reduces the duration of the SSD to some extent. In this regime, competing dynamical frequencies associated with near-resonant coupling and detuning generate a constructive interference in the atomic coherences, leading to a more efficient preservation of the spin--spin correlators responsible for steering.

In the intermediate region, for larger detuning values ($\Delta = \pm 2$), steering sudden death is eliminated further, SSD length becomes almost half, compared to the resonant case. Though, steerability is increased in the system, there is still a significant SSD present in the dynamics. However, the atom-atom subsystem becomes fully steerable in the dispersive regime of the detuned Hamiltonian throughout the dynamics as shown in the blue curve in the plot for $\Delta = \pm 5$. This revival can be attributed to the effective dispersive regime, in which real excitation exchange between atoms and cavities is strongly suppressed. This dispersive regime has been proven to be very practical and useful for some quantum information and computation purposes\cite{joshi2017qubit, wang2024dispersive, RevModPhys.93.025005}. In the dispersive regime, virtual rather than real photon exchange dominates: from Eq.~(30), the transfer amplitude $\beta_j(t) \propto g/\Omega_j$ becomes small for $\Delta_j \gg g$, so population transfer between the atom and cavity is strongly suppressed and the atomic state remains largely undisturbed. However, the eigenfrequency $\Omega_j = \sqrt{g^2+(\Delta_j/2)^2} \approx \Delta_j/2$ grows with detuning, so the phase $\alpha_j(t)$ in Eq.~(29) accumulates rapidly even though $\beta_j(t)$ stays small. Because $S_{A\rightarrow B}(t)$ depends directly on the off-diagonal coherence $\rho_{23}$ (Eq.~16), which tracks this phase accumulation, steering exhibits rapid oscillations at large $\Delta$. As a result, quantum coherence remains largely confined within the atomic subsystem, stabilizing the reduced atomic state against loss of steerability.

These findings demonstrate that detuning acts as a powerful control parameter for steering dynamics, inducing a monotonic transition from steering death to steering stabilization.

\section{Effects of dipole--dipole interaction on quantum steering}\label{sec5}

In this section, the effects of dipole-dipole interaction on the steerability dynamics are discussed. We consider two identical two-level atoms each locally coupled to a single-mode cavity field, and interacting with each other via a resonant dipole--dipole interaction (DDI). The influence of dipole-dipole interactions on atom--field dynamics has attracted considerable attention over the years due to its fundamental importance in cavity QED and quantum information processing \cite{laha2023dynamics,Khouja_2023,Evseev_2017}. In particular, recently, Khouja et. al\cite{Khouja_2023} has shown that addition of DDI in the system enhances the bidirectionality and stability of the steering dynamics. In  Ref.\cite{PhysRevA.44.2135}, Joshi et. al, have discussed the detailed dynamics of the two-atom system where two atoms are interacting with each other via dipole-dipole coupling. Evseev et. al., \cite{Evseev_2017}, have investigated the entanglement dynamics between two qubits in a non-resonant DJCM taking into account the direct dipole-dipole interaction between the qubits. The results show that the dipole-dipole parameter has great impact on the entanglement dynamics. In another recent paper, Mandal\cite{Mandal_2025} (one of the authors of this work), has shown that DDI helps removing the ESDs from the atom-atom and field-field entanglement dynamics.

Motivated by these findings, we now explore how DDI affects the quantum steering dynamics in the present model and whether it can serve as a controllable mechanism for protecting quantum correlations against decoherence.

The total Hamiltonian with the dipole interaction between the atoms can be expressed as
\begin{equation}
\hat{H} = \hat{H}_0 + \hat{H}_{\mathrm{I}} + \hat{H}_{\mathrm{dd}},
\end{equation}
with
\begin{align}
\hat{H}_0 &= \omega_a \left( \hat{\sigma}_A^+ \hat{\sigma}_A^- + \hat{\sigma}_B^+ \hat{\sigma}_B^- \right) 
+ \omega_c \left( \hat{a}^\dagger \hat{a} + \hat{b}^\dagger \hat{b} \right), \\
\hat{H}_{\mathrm{I}} &= 
g \left( \hat{\sigma}_A^+ \hat{a} + \hat{\sigma}_A^- \hat{a}^\dagger \right) 
+ g \left( \hat{\sigma}_B^+ \hat{b} + \hat{\sigma}_B^- \hat{b}^\dagger \right), \\
\hat{H}_{\mathrm{dd}} &= 
g_{\mathrm{dd}} \left( \hat{\sigma}_A^+ \hat{\sigma}_B^- + \hat{\sigma}_A^- \hat{\sigma}_B^+ \right).
\end{align}
Here $g$ is the atom--cavity coupling strength, and $g_{\mathrm{dd}}$ denotes the dipole--dipole interaction strength. The total excitation number operator
\begin{equation}
\hat{N} =
\hat{\sigma}_A^+ \hat{\sigma}_A^- + \hat{\sigma}_B^+ \hat{\sigma}_B^-
+ \hat{a}^\dagger \hat{a} + \hat{b}^\dagger \hat{b}.
\end{equation}
commutes with $\hat{H}$, implying that the dynamics preserves excitation number. We therefore restrict our analysis to the single-excitation subspace, spanned by
\begin{equation}
\mathcal{H}_{1} =
\left\{
\ket{eg,0,0},\;
\ket{ge,0,0},\;
\ket{gg,1,0},\;
\ket{gg,0,1}
\right\}.
\end{equation}
It is convenient to introduce symmetric and antisymmetric combinations of the atomic states,
\begin{equation}
\ket{\pm} =
\frac{1}{\sqrt{2}}
\left( \ket{eg} \pm \ket{ge} \right),
\end{equation}
as well as symmetric and antisymmetric photonic states,
\begin{equation}
\ket{G_\pm} =
\frac{1}{\sqrt{2}}
\left( \ket{gg,1,0} \pm \ket{gg,0,1} \right).
\end{equation}
This change of basis block-diagonalizes the Hamiltonian in the one-excitation sector. In the one-excitation subspace, all basis states have the same bare energy $\omega$ under $H_0$,
\begin{equation}
\hat{H}_0 \ket{\pm} = \omega \ket{\pm}, \quad
\hat{H}_0 \ket{G_\pm} = \omega \ket{G_\pm}.
\end{equation}
We therefore move to the interaction picture with respect to $H_0$, where the effective Hamiltonian is
\begin{equation}
\hat{H}_\mathrm{I} = e^{i H_0 t} (\hat{H} - \hat{H}_0) e^{-i H_0 t}
= \hat{H}_{\mathrm{JC}} + \hat{H}_{\mathrm{dd}}.
\end{equation}
Under the resonant condition and within the rotating-wave approximation, $H_I$ is time independent.

The dipole--dipole term acts solely within the atomic subspace and yields
\begin{equation}
\hat{H}_{\mathrm{dd}} \ket{\pm} = \pm g_{\mathrm{dd}} \ket{\pm}.
\end{equation}
Thus, the symmetric and antisymmetric atomic states acquire opposite energy shifts $\pm g_{\mathrm{dd}}$, corresponding to superradiant and subradiant channels, respectively.

The Jaynes--Cummings interaction couples atomic and photonic excitations as
\begin{align}
\hat{H}_{\mathrm{JC}} \ket{eg,0,0} &= g \ket{gg,1,0}, \\
\hat{H}_{\mathrm{JC}} \ket{ge,0,0} &= g \ket{gg,0,1}.
\end{align}
Forming symmetric and antisymmetric combinations, we obtain
\begin{align}
\hat{H}_{\mathrm{JC}} \ket{+} &= \sqrt{2} g \ket{G_+}, \\
\hat{H}_{\mathrm{JC}} \ket{-} &= \sqrt{2} g \ket{G_-}.
\end{align}
No coupling occurs between states of opposite symmetry, and the Hamiltonian decomposes into two independent subspaces.

In the basis $\{ \ket{\pm}, \ket{G_\pm} \}$, the interaction-picture Hamiltonian takes the block-diagonal form
\begin{equation}
\hat{H}_\mathrm{I} = \hat{H}_+ \oplus \hat{H}_-,
\end{equation}
with
\begin{equation}
H_\pm =
\begin{pmatrix}
\pm g_{\mathrm{dd}} & \sqrt{2} g \\
\sqrt{2} g & 0
\end{pmatrix}.
\label{eq:Hpm}
\end{equation}
Each block describes an effective two-level system with detuning $\pm g_{\mathrm{dd}}$ and coupling strength $\sqrt{2} g$.

Diagonalizing $H_\pm$ yields the eigenvalues
\begin{equation}
\lambda_{\pm}^{(1,2)} =
\frac{1}{2}
\left(
\pm g_{\mathrm{dd}} \pm \sqrt{g_{\mathrm{dd}}^2 + 8 g^2}
\right).
\end{equation}
The energy splitting between the dressed states defines the generalized Rabi frequency
\begin{equation}
\boxed{
\Omega = \sqrt{g_{\mathrm{dd}}^2 + 8 g^2}.
}
\end{equation}
Thus, the dipole--dipole interaction acts as an effective detuning that modifies the Rabi oscillations of the coupled atom--cavity system.

Assuming the cavities are initially in the vacuum state and the atomic state is
\begin{equation}
\ket{\psi(0)} = c_+(0) \ket{+} + c_-(0) \ket{-},
\end{equation}
the Schrödinger equation in each symmetry sector yields
\begin{equation}
c_\pm(t) =
c_\pm(0)
\left[
\cos\!\left( \frac{\Omega t}{2} \right)
- i \frac{\pm g_{\mathrm{dd}}}{\Omega}
\sin\!\left( \frac{\Omega t}{2} \right)
\right].
\end{equation}
Due to excitation-number conservation, the dynamics induced by the dipole--dipole interaction can be treated analytically within the single-excitation sector. By diagonalizing the effective Hamiltonian in this subspace, we obtain closed-form expressions for the atomic probability amplitudes and coherences. Although the resulting steering parameter involves nonlinear functions of the reduced density matrix and must be evaluated numerically, the analytical structure of the dynamics reveals the physical mechanism by which dipole--dipole interaction suppresses steering sudden death. This exact analytical solution explicitly demonstrates how the dipole--dipole interaction splits the symmetric and antisymmetric channels and modifies their coherent dynamics.

The above decomposition shows that the dipole--dipole interaction introduces opposite detunings in the symmetric and antisymmetric sectors, leading to distinct oscillation frequencies and coherence lifetimes. Since $S_{A\to B}(t)$ depends on the atomic coherence $\rho_{23}(t)=\langle eg|\hat\rho_{AB}(t)|ge\rangle$ through $T_{xx}, T_{yy}$ in Eq.~(18), the two-frequency interference between the $\lambda_\pm$ (Eq.~48) prevents $\rho_{23}(t)$ from vanishing at the instants where the single-frequency, non-interacting case would give SSD.

\begin{figure}
    \centering
    \includegraphics[width=\linewidth]{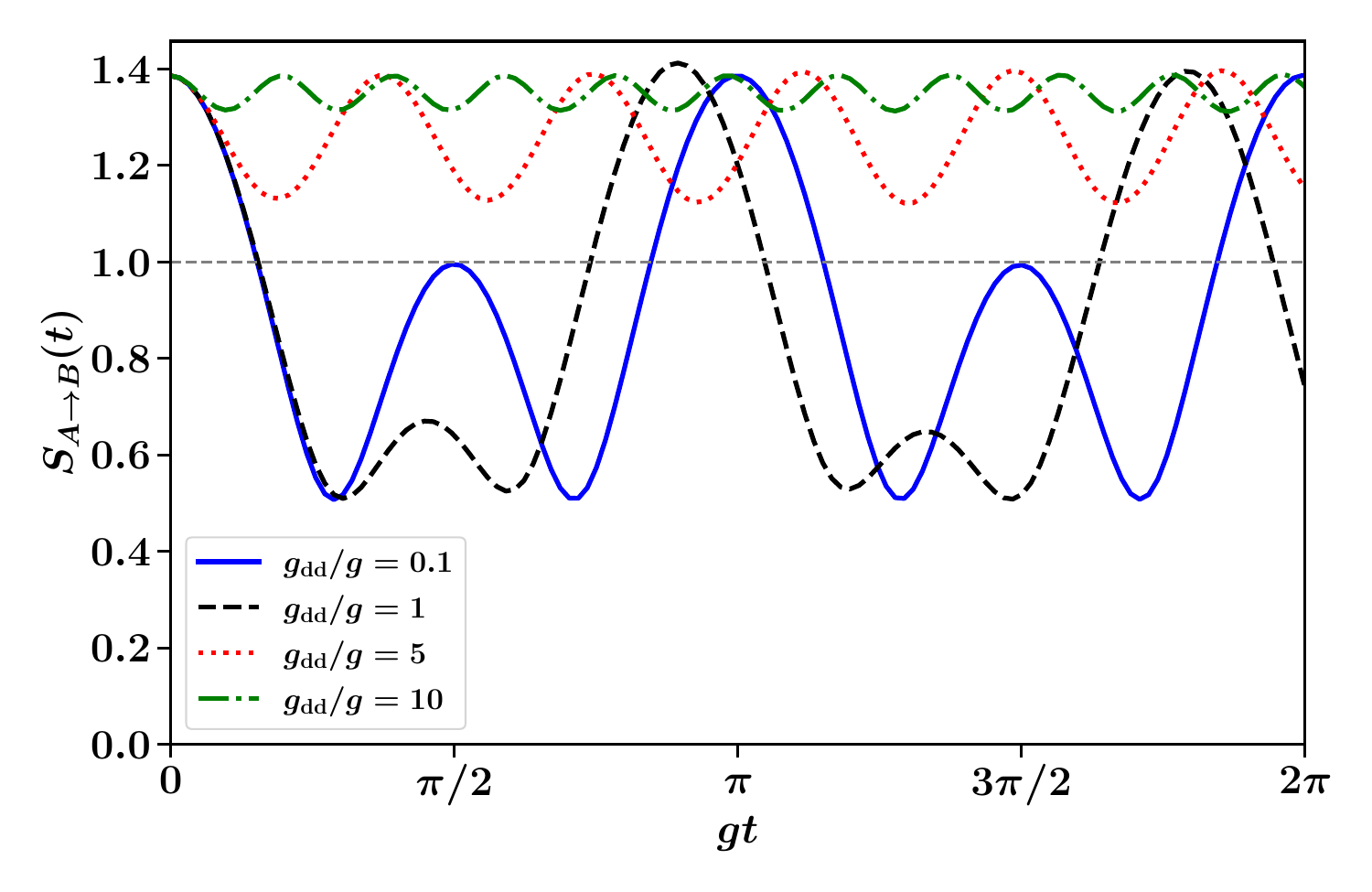}
    \caption{Dynamics of quantum steering in the presence of dipole--dipole interaction for the initial Werner-state parameter $p=0.8$. The steering behavior is analyzed for different values of the normalized dipole--dipole coupling strength $g_\mathrm{dd}/g$, ranging from $0.1$ to $10$, illustrating the transition from weak to strong interaction regimes.}
    \label{dipole_int_steer}
\end{figure}

The dipole--dipole coupling strength $g_{\mathrm{dd}}$ is treated as an independent, tunable parameter. Values $g_{\mathrm{dd}}/g \gtrsim 1$ naturally arise in platforms such as circuit QED, Rydberg-atom systems, and waveguide-mediated interactions, where direct inter-emitter coupling can dominate over atom--field exchange. 

Figure \ref{dipole_int_steer} shows the time evolution of the atomic steering $S_{A\to B}(t)$ for different strengths of dipole--dipole interaction $g_{\mathrm{dd}}/g$, with fixed visibility $p=0.8$. In the weak-coupling regime, steering exhibits pronounced sudden death intervals, during which $S_{A\to B}(t)$ falls below 0.6. As the interaction strength increases, these intervals progressively shrink and eventually disappear. For sufficiently strong dipole--dipole interaction e.g, for $g_{\mathrm{dd}}/g \ge 5$, steering remains finite at all times though oscillatory, indicating complete suppression of steering sudden death.

\begin{figure}[ht!]
\centering
\includegraphics[width=\linewidth]{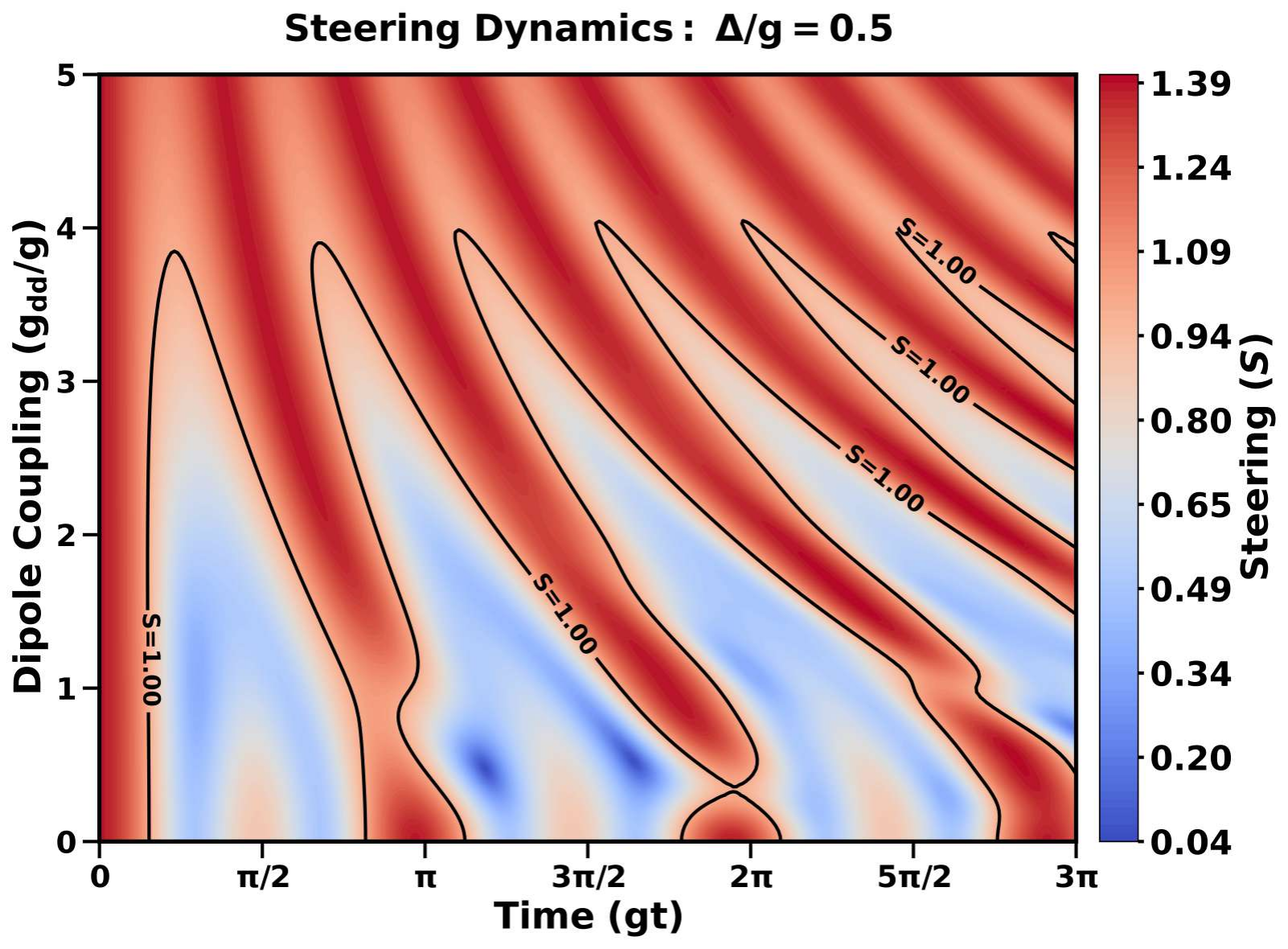}
\caption{Contour plot of the quantum steering dynamics as a function of the scaled time $gt$ and normalized dipole--dipole coupling strength $g_\mathrm{dd}/g$ for a fixed detuning $\Delta/g=\pm\,0.5$ in the near-resonant regime. The initial Werner-state parameter is chosen as $p=0.8$. The plot illustrates the steering behavior when dipole--dipole interaction and atom--cavity detuning are simultaneously present in the system.}
\label{dipole_det_0.5}
\end{figure}

This behavior arises because the dipole-dipole interaction drives a coherent excitation exchange between the atoms at frequency $g_{dd}$. When $g_{dd} \gg g$, this direct interatomic coupling dominates over the atom-cavity coupling $g$, effectively detuning the atomic states from the cavity modes and suppressing photon emission. Because the CJWR steering parameter relies on transverse spin correlations—which are proportional to the off-diagonal atomic coherence—this energy shift keeps the state within the two-atom subspace. As a result, the off-diagonal coherence remains large, preventing the steering parameter from dropping below the classical threshold and thus eliminating steering sudden death.

\section{Effects of atom-cavity detuning and dipole--dipole interaction on quantum steering}\label{sec5b}
\begin{figure}[ht!]
\centering
\includegraphics[width=\linewidth]{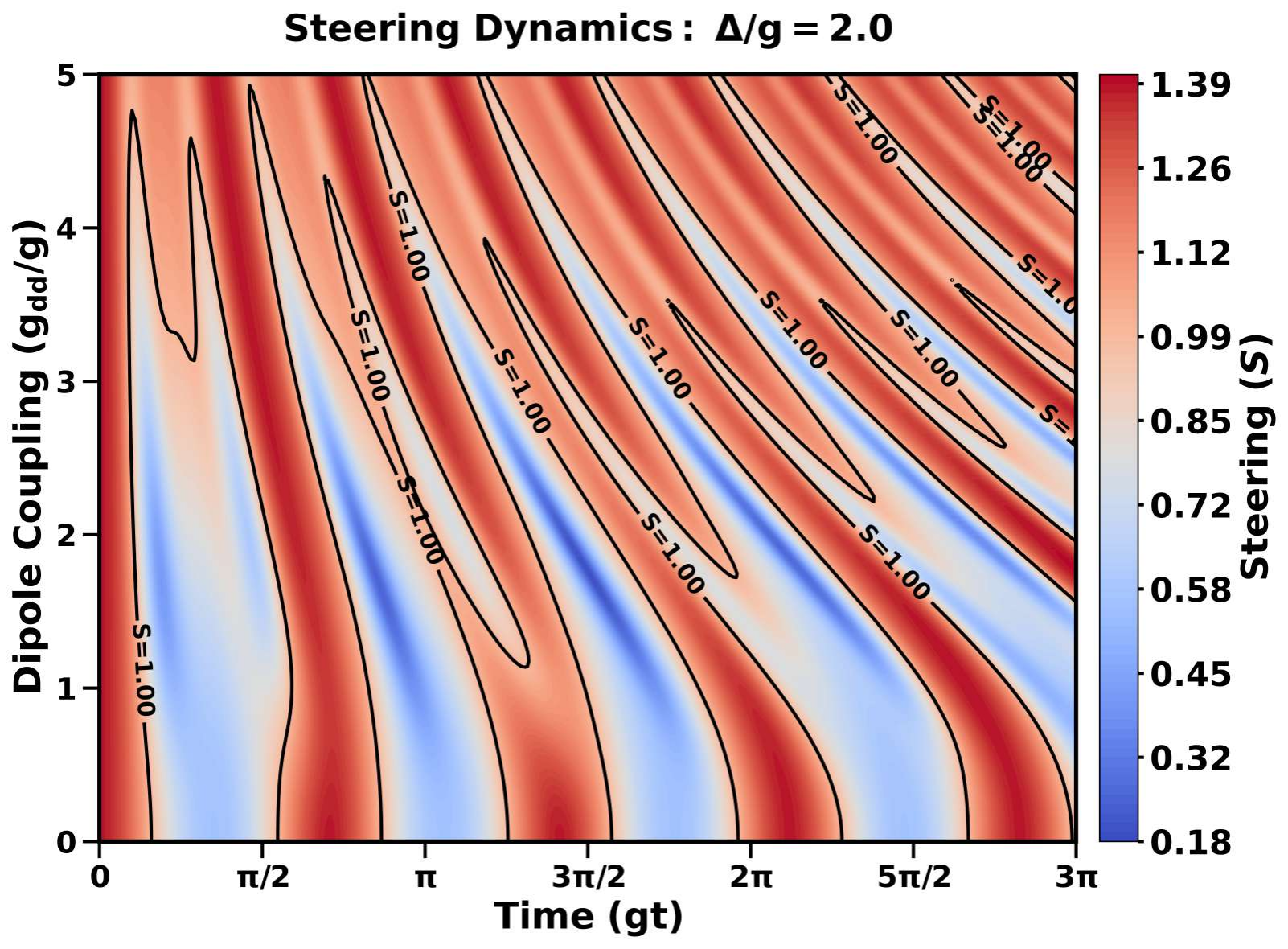}
\caption{Same as Fig.~\ref{dipole_det_0.5}, but for the intermediate detuned regime with $\Delta/g = \pm \,2$.  For a larger detuning, the regions of steerability have decreased.}
\label{dipole_det_2}
\end{figure}

We further extend our analysis by incorporating dipole--dipole interactions in the presence of atomic detuning, with the aim of identifying parameter regimes where these two effects can be optimally combined to control the steering dynamics. The behavior of the system for varying dipole-dipole coupling strength $g_{\mathrm{dd}}$, at fixed detuning $\Delta$, is illustrated in Figs.~\ref{dipole_det_0.5} and \ref{dipole_det_2}.

From Fig.~\ref{dipole_det_0.5}, it is evident that beyond a certain threshold of $g_{\mathrm{dd}}$, steering sudden death (SSD) re-emerges in the dynamics, despite being absent when only dipole--dipole interaction is present. As the detuning is increased further ($\Delta = \pm 2$), SSD becomes more pronounced, as shown in Fig.~\ref{dipole_det_2}. This behavior can be understood from the interplay between detuning and dipole--dipole interaction. In the detuned regime, the effective atomic transition frequency shifts such that it becomes resonant with the energy scale associated with the dipole--dipole coupling. As a consequence, the atom--field interaction, which was effectively suppressed by the dipole--dipole interaction alone, is partially restored.

This reactivation of atom-field coupling induces intrinsic local decoherence in the system, leading to the reappearance of steering sudden death. Therefore, while both detuning and dipole--dipole interaction independently act as mechanisms to suppress SSD, their combined effect can counterintuitively reintroduce decoherence pathways that degrade steerability. From these observations, we conclude that detuning and dipole--dipole interaction must be used with care to control steerability. They may be more effective in mitigating SSD when employed individually, rather than simultaneously, in the parameter regimes considered.

\section{Conclusion}
\label{sec6}
In this work, we have investigated the dynamical behavior of quantum steering between two atoms in a double Jaynes--Cummings model, with particular emphasis on the state of steering sudden death under the inclusion of detuning and interatomic interactions. Our analysis combines analytical insight based on Werner-type initial states with numerical simulations, allowing us to understand the physical mechanisms responsible for the observed stabilization of steerability. We identify non-analytic transitions in the fidelity dynamics at the boundaries of steering collapse and revival. These features arise from the rapid restructuring of quantum correlations as the atomic state traverses the Local Hidden State (LHS) manifold. The synchronization of these `cusp'-like features along with the steering threshold indicates that fidelity dynamics may serve as a sensitive diagnostic signature for the onset of SSD. 

A central result of our study is that the inclusion of detuning and dipole--dipole coupling individually can completely remove steering sudden death from the atom-atom subsystem, even in regimes where steering would otherwise vanish abruptly in the noninteracting model. From a physical perspective, the suppression of steering sudden death can be attributed to the role of interactions and detuning in protecting atomic coherences against decoherence induced by atom--cavity coupling. In the absence of interactions, the exchange of excitations with the cavity fields leads to a rapid degradation of the conditional correlations required for steering, resulting in finite-time disentanglement and steering death. On the other hand, the dipole-dipole interaction introduces additional coherent pathways that dynamically stabilize the non-local atomic correlation. Another interesting result is that detuning and dipole--dipole interaction, while individually suppressing steering sudden death, can jointly restore decoherence pathways and reintroduce SSD, 
 
Our key result is that through control parameters, quantum correlations can be restored in the DJCM. This establishes steering as a controllable dynamical feature rather than a static property.
Our study shows that interaction-induced stabilization of steering does not necessarily imply enhanced Bell nonlocality or entanglement, reinforcing the view that steering captures a unique aspect of nonclassicality. In this sense, the removal of steering sudden death reflects the preservation of operationally accessible quantum correlations rather than a simple increase in total correlation content.

Finally, we emphasize that the effects reported here are robust across a wide range of system parameters, including visibility, interaction strength, and detuning. These features make the double Jaynes--Cummings model a versatile platform for exploring controlled steering dynamics in multipartite open quantum systems. Our results provide new insight into how dipole-dipole interactions can be harnessed to protect and manipulate quantum correlations, and open avenues for further studies of steering flow, control, and resource use in complex quantum systems.

\begin{acknowledgments}
The authors gratefully acknowledge support from the Discovery Grant program of the Natural Sciences and Engineering Research Council of Canada (NSERC) and the NSERC Chairs for Women in Science and Engineering program.Funding reference number RGPIN-2019-06387 (CR).  We acknowledge computational resources from the Digital Research Alliance of Canada. Wilfrid Laurier
University is located in the traditional territory of the
Neutral, Anishnawbe and Haudenosaunee peoples. We
thank them for allowing us to conduct this research on
their land. The authors also acknowledge the use of Google's Gemini AI for language editing and assistance in improving the clarity and phrasing of the manuscript. All theoretical formulations, calculations, and physical interpretations were developed and verified solely by the authors.
\end{acknowledgments}

\appendix

\section{Analytical expression of the reduced density matrix}
\label{app_A}

The initial density matrix for the Werner state is expressed as
\begin{equation}
\rho_W(p) =
\begin{pmatrix}
\frac{1-p}{4} & 0 & 0 & 0 \\[6pt]
0 & \frac{1+p}{4} & -\frac{p}{2} & 0 \\[6pt]
0 & -\frac{p}{2} & \frac{1+p}{4} & 0 \\[6pt]
0 & 0 & 0 & \frac{1-p}{4}
\end{pmatrix}.
\end{equation}

To properly characterize the quantum correlations between two atoms ($A$ and $B$) interacting with their respective cavities ($a$ and $b$), we derive the reduced density matrix $\hat{\rho}_{AB}(t)$ by tracing out the cavity degrees of freedom. We define $C \equiv \cos(gt)$ and $S \equiv \sin(gt)$.

The resonant Double Jaynes-Cummings Hamiltonian leads to the following evolution for the atomic basis states when the cavities are initially in the vacuum state $|00\rangle$:
\begin{align}
|gg,00\rangle &\to |gg,00\rangle ,\\
|eg,00\rangle &\to C|eg,00\rangle - iS|gg,10\rangle, \\
|ge,00\rangle &\to C|ge,00\rangle - iS|gg,01\rangle, \\
|ee,00\rangle &\to C^2|ee,00\rangle - iCS|eg,01\rangle \nonumber \\
&\quad - iCS|ge,10\rangle - S^2|gg,11\rangle.
\end{align}

Taking the partial trace over the cavity modes $\text{Tr}_{ab}$ for each state results in the following atomic density operators $\hat{\rho}_\alpha(t)$:
\begin{itemize}
    \item From $|gg\rangle$: $\hat{\rho}_{gg}(t) = |gg\rangle\langle gg|$
    \item From $|eg\rangle$: $\hat{\rho}_{eg}(t) = C^2|eg\rangle\langle eg| + S^2|gg\rangle\langle gg|$
    \item From $|ge\rangle$: $\hat{\rho}_{ge}(t) = C^2|ge\rangle\langle ge| + S^2|gg\rangle\langle gg|$
    \item From $|ee\rangle$: $\hat{\rho}_{ee}(t) = C^4|ee\rangle\langle ee| + C^2S^2|eg\rangle\langle eg| + C^2S^2|ge\rangle\langle ge| + S^4|gg\rangle\langle gg|$
\end{itemize}

The initial state is a Werner state $\hat{\rho}_{AB}(0) = p|\Psi^-\rangle\langle \Psi^-| + \frac{1-p}{4}\hat{\mathbb{I}}_4$.

\textit{Bell Component.}---The state $|\Psi^-\rangle = \frac{1}{\sqrt{2}}(|eg\rangle - |ge\rangle)$ evolves into:
\begin{equation}
\hat{\rho}_{AB}^{(\Psi)}(t) = C^2 |\Psi^-\rangle\langle \Psi^-| + S^2 |gg\rangle\langle gg|.
\end{equation}

\textit{Mixed Component.}---The noise part is the sum of the basis state evolutions:
\begin{equation}
\hat{\rho}_{AB}^{(\text{noise})}(t) = \frac{1-p}{4} \left[ \hat{\rho}_{ee}(t) + \hat{\rho}_{eg}(t) + \hat{\rho}_{ge}(t) + \hat{\rho}_{gg}(t) \right].
\end{equation}

Combining both parts, the total reduced density matrix $\hat{\rho}_{AB}(t)$ in the basis $\{|ee\rangle, |eg\rangle, |ge\rangle, |gg\rangle\}$ is an $X$-state:
\begin{equation}
\hat{\rho}_{AB}(t) = 
\begin{pmatrix}
\rho_{11} & 0 & 0 & 0 \\
0 & \rho_{22} & \rho_{23} & 0 \\
0 & \rho_{32} & \rho_{33} & 0 \\
0 & 0 & 0 & \rho_{44}
\end{pmatrix},
\end{equation}
where the non-zero matrix elements are:
\begin{align}
\rho_{11} &= \frac{1-p}{4}C^4, \\
\rho_{22} = \rho_{33} &= \frac{C^2}{4}\left[ 2p + (1-p)(1+S^2) \right], \\
\rho_{23} = \rho_{32} &= -\frac{p}{2}C^2, \\
\rho_{44} &= pS^2 + \frac{1-p}{4}(1+S^2)^2.
\end{align}

In matrix form,
\begin{widetext}
\begin{equation}
\rho_{AB}(t) =
\begin{pmatrix}
\frac{1-p}{4} C^4 & 0 & 0 & 0 \\[6pt]
0 & \frac{C^2}{4}\left[2p + (1-p)(1+S^2)\right] & -\frac{p}{2} C^2 & 0 \\[6pt]
0 & -\frac{p}{2} C^2 & \frac{C^2}{4}\left[2p + (1-p)(1+S^2)\right] & 0 \\[6pt]
0 & 0 & 0 & p S^2 + \frac{1-p}{4}(1+S^2)^2
\end{pmatrix}.
\end{equation}
\end{widetext}

\section{Analytical expression of fidelity in DJCM}
\label{app_B}

\noindent \textbf{The Diagonal Elements:}\\

The states $|ee\rangle$ and $|gg\rangle$ form a diagonal subspace. The initial populations are:
\begin{equation}
\rho_{11}(0) = \frac{1-p}{4}, \quad \rho_{44}(0) = \frac{1-p}{4}.
\end{equation}
The time-evolved populations, including the feed-down from the $|ee\rangle$ state, are:
\begin{align}
\rho_{11}(t) &= \frac{1-p}{4}\cos^4(gt), \\
\rho_{44}(t) &= p\sin^2(gt) + \frac{1-p}{4}(1+\sin^2(gt))^2.
\end{align}
Since $\rho_{11}$ and $\rho_{44}$ are already diagonal in this subspace, the classical 
fidelity formula gives the contribution to $\sqrt{F}$ as
\begin{equation}
\sqrt{F_{\text{diag}}} = \sqrt{\rho_{11}(0)\rho_{11}(t)} + \sqrt{\rho_{44}(0)\rho_{44}(t)}.
\end{equation}

\noindent \textbf{The Central $2 \times 2$ Block:}\\

The central block acts on the subspace $\{|eg\rangle, |ge\rangle\}$. Both $\rho(0)$ and $\rho(t)$ are symmetric in this basis, taking the form $\begin{pmatrix} a & b \\ b & a \end{pmatrix}$. Such matrices are diagonalized in the Bell basis $\{|\Psi^+\rangle, |\Psi^-\rangle\}$ with eigenvalues $\lambda_{\pm} = a \pm b$.\\

\noindent{\textbf{Initial Eigenvalues} ($t=0$):}\\

With $\rho_{22}(0) = \frac{1+p}{4}$ and $\rho_{23}(0) = -\frac{p}{2}$:
\begin{align}
\lambda_+(0) &= \frac{1+p}{4} - \frac{p}{2} = \frac{1-p}{4} \\
\lambda_-(0) &= \frac{1+p}{4} + \frac{p}{2} = \frac{1+3p}{4}
\end{align}
\\
\textit{Time-evolved eigenvalues.}---With $\rho_{22}(t) = \frac{\cos^2(gt)}{4}[2p + (1-p)(1+\sin^2(gt))]$ and $\rho_{23}(t) = -\frac{p}{2}\cos^2(gt)$, the eigenvalues are given by:
\begin{align}
\lambda_+(t) &= \rho_{22}(t) + \rho_{23}(t) \nonumber \\
&= \frac{1-p}{4}\cos^2(gt)(1+\sin^2(gt)), \\
\lambda_-(t) &= \rho_{22}(t) - \rho_{23}(t) \nonumber \\
&= \frac{\cos^2(gt)}{4}\left[ 4p + (1-p)(1+\sin^2(gt)) \right].
\end{align}

Applying the fidelity formula above to the eigenvalues $\lambda_\pm(0)$ and 
$\lambda_\pm(t)$ derived in Eqs.~(B5)--(B8), the contribution to $\sqrt{F}$ from this 
subspace is
\begin{equation}
\sqrt{F_{\text{block}}} = \sqrt{\lambda_+(0)\lambda_+(t)} + \sqrt{\lambda_-(0)\lambda_-(t)}.
\end{equation}

\textit{Total fidelity expression.}---The total square root of the fidelity is the sum of the square roots of the overlaps of all four eigenvalues:
\begin{align}
\sqrt{F(t)} ={}& \sqrt{\rho_{11}(0)\rho_{11}(t)} + \sqrt{\rho_{44}(0)\rho_{44}(t)} \nonumber \\
& + \sqrt{\lambda_+(0)\lambda_+(t)} + \sqrt{\lambda_-(0)\lambda_-(t)}.
\end{align}

Substituting all derived terms:
\begin{align}
\sqrt{F(t)} &= \frac{1-p}{4}\cos^2(gt) \notag \\
&+ \frac{\sqrt{1-p}}{2}\sqrt{p\sin^2(gt) + \frac{1-p}{4}(1+\sin^2(gt))^2} \notag \\
&+ \frac{1-p}{4}\cos(gt)\sqrt{1+\sin^2(gt)} \notag \\
&+ \frac{\cos(gt)}{4}\sqrt{(1+3p)[4p + (1-p)(1+\sin^2(gt))]}.
\end{align}
The final fidelity is obtained by squaring:
\begin{equation}
F(t) = \left( \sqrt{F_{\text{diag}}} + \sqrt{F_{\text{block}}} \right)^2.
\end{equation}

\bibliographystyle{naturemag}
\bibliography{ref_steer}

\end{document}